# Science Use in Regulatory Impact Analysis: The Effects of Political Attention and Controversy

Mia Costa[*,1], Bruce A. Desmarais [†,2], and John A. Hird [‡,1,3]

[1]Department of Political Science, University of Massachusetts Amherst
[1]Department of Political Science, The Pennsylvania State University
[3]College of Social and Behavioral Sciences, University of Massachusetts Amherst

**Abstract**

Scholars, policymakers, and research sponsors have long sought to understand the conditions under which scientific research is used in the policymaking process. Recent research has identified a resource that can be used to trace the use of science across time and many policy domains. US federal agencies are mandated by executive order to justify all economically significant regulations by regulatory impact analyses (RIAs), in which they present evidence of the scientific underpinnings and consequences of the proposed rule. To gain new insight into when and how regulators invoke science in their policy justifications, we ask: does the political attention and controversy surrounding a regulation affect the extent to which science is utilized in RIAs? We examine scientific citation activity in all 101 economically significant RIAs from 2008-2012 and evaluate the effects of attention – from the public, policy elites and the media – on the degree of science use in RIAs. Our main finding is that regulators draw more heavily on scientific research when justifying rules subject to a high degree of attention from outside actors. These findings suggest that scientific research plays an important role in the justification of regulations, especially those that are highly salient to the public and other policy actors.

[*]micosta@polsci.umass.edu
[†]bdesmarais@psu.edu
[‡] jhird@sbs.umass.edu

# 1   Introduction

Perceived accountability and transparency in regulatory policymaking depends in large part on the justification and use of evidence by agency officials to develop legitimate policies. Without an electoral mandate, regulators must rely on domain expertise in policymaking to garner political support from various actors including elected officials, the public, industry, and advocacy groups (Turnpenny et al., 2009). Agencies are not only expected but also required to base their regulatory justifications on a solid scientific basis, requirements that were first mandated by Presidential Executive Order 12291 in 1981. While relevant science has long been crucial to sound regulatory decisions, agencies' reliance on technical expertise is becoming increasingly significant in the current political environment as the use of scientific evidence is "politicized." For example, as the Obama administration takes up issues such as climate change and public health, the issue of science is also brought into the political spotlight. Remarking in 2009 on a White House directive, President Barack Obama emphasized the prioritization of science in public policy by promising "to ensure that in this new Administration, we base our public policies on the soundest science; that we appoint scientific advisors based on their credentials and experience, not their politics or ideology" (The White House, 2009). Whether shifts in the executive actually stimulate change in agencies' use of science is an open question outside the scope of this paper. However, presidential pronouncements such as this illustrate the potential for science to help inform inherently political decisions, which also suggests that political attention may lead to greater use of science.

Despite the recent prioritization of research in policymaking, there is surprisingly little systematic evidence regarding the relationships between science in regulations and the conditions under which research is invoked. After a thorough review of the literature, a recent US National Research Council (NRC) report notes that "...despite their considerable value in other respects, studies of knowledge utilization have not advanced understanding of the use of evidence in the policy process much beyond the decades-old [1978] National Research Council report" (NRC, 2012, p. 51). Past scholarship on knowledge utilization in public policy often focused on the relationship between research findings and specific policy outcomes, yet strong linkages were rarely uncovered (Contandriopoulos et al., 2010). Instead, research moved to understand the multiple ways that research engages with policymaking, besides direct impact on policymaking outcomes (Weiss, 1979). While numerous studies pointed to ways research interacted with policymaking, most research findings have been limited to particular policy areas or policymaking episodes (Whiteman, 1985; Farrow, 1991; Bogenschneider et al., 2000; Rich, 2004, e.g.). Indeed, the NRC report concludes: "the scholarship on use to date is inadequate" (NRC, 2012, p.33).

Notwithstanding the lack of broadly generalizable findings regarding knowledge utilization in public policy, several scholars hold that political support for regulations relies heavily on the quality of the evidence and arguments supporting the agency's justifications for proposed rules (see, e.g., Lupia and McCubbins 1994a; Majone 2002; Vibert 2007). When faced with a highly salient rulemaking task, we hypothesize that regulators will draw more heavily upon scientific evidence in order to head off opposing arguments either in the formation of the regulation or in any possible subsequent lawsuits. Regulatory impact analyses (RIAs) in the U.S. are the public expression of such justifications and represent an unusually coherent policy argument from the regulators' perspectives. "RIA is a particularly fascinating case for the analysis of the role of knowledge in policy-making because it has quasi-scientific ambitions, but also takes place at the heart of government where political decisions are transformed into laws, regulations and other policy instruments" (Hertin et al., 2009, 413). Since the Executive Order 12291 in 1981, all rules estimated to have an economic impact of $100 million or more are determined to be "economically significant" by the Office of Information and Regulatory Affairs (OIRA) in the US Office of Management and Budget (OMB). The author agencies must then submit an RIA to OIRA for review and approval before the promulgation of the regulation.

RIA has opened up a new line of research for the field of knowledge utilization. Some research, for example, has demonstrated that regulators rely on scientific research to assess benefits and costs of proposed regulations, although the quality of RIAs varies widely across agencies (Ellig, McLaughlin and Morrall III, 2013). Recent research by Desmarais and Hird (2013) shows that for US federal regulations, RIAs are more likely to invoke research published in the most prominent scholarly journals. Thus, initial findings suggest that the invocation of science is not mere window dressing on regulations, but appears overall to be a purposeful use of high quality science to bolster arguments supporting regulatory proposals. We expand upon this research by broadening the scope to include an analysis of the relationship between the surrounding political environment and invocation of science in RIAs across a wide range of policy domains.

There are several reasons why agencies would invoke science in their regulatory justifications. The fact that policymakers reference science does not necessarily imply that the science influenced the formation of a policy proposal. Indeed, past research has found that science is often invoked to validate an existing policy decision (Hird, 2005, ch.2). In such instrumental cases of use, science constrains the set of policy options to those that can be justified through the scientific literature, but does not exercise an active influence on policy development. Regardless of where an instance of science use stands on the spectrum from highly influential to purely instrumental, regulators draw on a wide range of scientific evidence to establish their expertise in specific areas regarding public and environmental protection (Doremus, 2007). Further, it is in the interest of officials, elected or appointed, to understand the potential consequences of proposed policies, and using the best possible research improves predictions of projected outcomes (Lupia and McCubbins, 1994b). Finally, scientific support for rulemakings helps agencies with political and legal challenges, both prior and subsequent to issuing the regulation. Jasanoff (1990), for example, shows that regulations issued by agencies like the Environmental Protection Agency are regularly contested in court, which has motivated agencies to bolster their rule justification from the outset. For all of these reasons it is likely that agencies take care to use science for all rulemakings, but the amount of science used will likely vary depending on the agency, issue, or external conditions surrounding the rule.

An open question, of both intellectual and practical import, is whether political attention and controversy surrounding regulatory policymaking affects the nature and volume of science invoked in justifying policy decisions. Reviewing the literature on research utilization, Weiss (1979) illustrates that there is a "diversity of perspectives" regarding how and when agencies use evidence in policy and suggests that the most common view, that agencies find or commission science to help solve specific policy dilemmas, is insufficient to explain the majority of cases of research utilization (30). One reason the problem-solving model cannot fully capture the process of research utilization is because strong political interests complicate the ways in which evidence enters the policy arena (Weiss, 1979). For one, if agencies anticipate political and legal challenges, then regulators may act strategically to bolster scientific citations for rules where the chances of such challenges are greater. Some studies have focused on the quality of evidence invoked in regulations (Ellig and McLaughlin, 2012; Ellig, McLaughlin and Morrall III, 2013) but we still know very little about the conditions that influence the evidence used. In this paper, we examine the relationship between the political attention surrounding a regulation measured across three dimensions (public comments, meetings with OIRA officials, and news coverage) and the level of science invoked by agency officials in their regulatory justifications. The central hypothesis we investigate is that regulators draw upon scientific research more heavily when creating a regulation that is subject to widespread political scrutiny. This is tested against the null hypothesis that agencies act without regard to external pressures; the use of science in RIAs remains unaffected when accounting for the volume of public comments, OIRA meetings, and news coverage.

US regulatory agencies have standard operating procedures (SOPs) that, to varying degrees, address the use of science in the policymaking process. As such, we would expect the baseline levels of science use to vary across agencies and exhibit within-agency consistencies, but these SOPs cannot explain the variability of science use within agencies, across different regulations. In contrast, a finding that agencies do indeed respond differently in their use of science based on external attention and controversy suggests that agencies are capable of invoking more or better science, conditional on the resources available. Such a finding would have important policy implications, since agencies may be able to acquire greater capabilities over time if so required. Since our measures of political attention largely precede the release of the RIA, we can be confident that political attention and controversy are driving agency use of science, not vice-versa. We test this relationship below using a novel dataset of citation counts in 101 RIAs between 2008- 2012. We find significant evidence that regulators draw more heavily on scientific research when justifying rules subject to a high degree of attention from outside actors.

## 2 Political Attention and Controversy

Since the passage of the Administrative Procedures Act (APA) in 1946, US federal law has provided mechanisms for the public to offer input formally throughout the rulemaking process (Shapiro, 2008). These input procedures play an essential role in assuring some level of representativeness in policies that are otherwise crafted by unelected bureaucrats. There are three origins of political attentiveness to which we would expect regulators to be consistently responsive. The first is that of the general public. Politically active citizens are responsible for the electoral conditions faced by regulators' political principals – both members of Congress and the President – that present a strong incentive for regulators to consider the level of general public attentiveness to a regulation. The second distinct source of attention is that allotted by elite policy actors (e.g., other federal officials, businesses and advocacy groups). These actors often represent the targets of legislation and are the most prominent stakeholders in the policymaking process. The third source of attention comes from the media. Through coverage of policies, the media hold the potential to set the public agenda regarding a proposed regulation and heavily influence the public's attention to the process.

Public comments on proposed regulations represent what are perhaps the best-known examples of public input to the regulatory process and are commonly studied as indicators of participatory rulemaking at work. Regulators publish proposed regulations in the Federal Register, at which point a notice-and-comment period begins. The public can submit comments on the proposed rule during a comment period that generally lasts between 30 to 90 days, depending on the complexity of the rule. Federal regulatory agencies are required through the APA to accept and respond to substantive comments prior to issuing final regulations. This process also includes the provision of, and often revisions to, regulatory impact analyses that justify agency decision-making for economically significant RIAs. Comments are consistently found to affect rule development (Naughton et al., 2009; Shapiro, 2008; Yackee, 2006; Yackee and Yackee, 2006), a finding that suggests public participation matters in the regulatory process. However, others note (Noveck, 2004; Romsdahl, 2005; Shulman, 2009; Zavestoski, Shulman and Schlosberg, 2006) that web-based commenting platforms open the floodgates to insubstantial comments, degenerating the process into a costly "notice-and-spam" routine. Nevertheless, in both large-*N* quantitative and small-*N* qualitative studies, researchers have found that regulators are responsive to both the quantity and content of comments (Balla and Daniels, 2007; Karpf, 2010; Naughton et al., 2009; Shapiro, 2008; Yackee, 2006; Yackee and Yackee, 2006). Naughton et al.'s (2009) interview with one agency official succinctly summarizes how seriously they consider both science and controversy in comments:

> "The relatively high value placed on hard data in comments is best summed up by the interviewee who stated, 'We look at every comment; we consider every comment. But unless there is [sic] data supporting the position, its just not that useful in the rulemaking process.' With respect to political controversy, one interviewee stated, 'it gives you the opportunity to sort of test the waters and see what people think" (270).

Public comments communicate attention through a widely accessible and low-cost mode of participation. Previous work shows that a high volume of comments from more influential parties, such as business interests, differentially affect rule development (Yackee and Yackee, 2006). Motivated by this finding, we look to the Office of Information and Regulatory Affairs (OIRA), within the Office of Management and Budget, which is responsible for White House oversight of rulemaking and has broad influence over regulatory policymaking. Moreover, outside actors can request meetings with OIRA to discuss regulatory policy issues, and OIRA is required to keep a public log of the participants in such meetings, as well as any supporting documents submitted (Croley, 2003). These meetings generally represent participation by organized interests, often business interests and trade associations in the Washington, DC area. Croley (2003) finds that more than 80 percent of meetings with OIRA involved what he classifies as narrow interests, with 56 percent involving only narrow interest participants. These meetings are much less accessible to the general public, as they require coordination and planning as well as considerable time commitment and location near or travel to Washington, DC.

Finally, we see attention from the mass media as related to, yet distinct from, both broad public attention and elite interests. News coverage both depends on and holds the potential to expand the salience of regulatory developments, especially as that salience is focused on regulators and not just the federal government at large. The media covers stories that may depend on pre-existing public attention, but can also drive the way in which a proposed policy is perceived in the public and elite circles. For example, Kraft and Vig (1984) note that media coverage played a significant role in discrediting many of President Reagan's early environmental policy initiatives. In a broader finding, Ruder (Forthcoming) shows that news attention to the executive branch can help move the attribution of responsibility for policy change away from Congress and towards the President and regulatory agencies.

In sum, the development of a final regulation relies on the pressures that influence agency officials' considerations in writing the rule. But while it is clear many studies consider outside interests central to the rulemaking process, they suffer from a number of shortcomings when it comes to drawing generalizable conclusions that span policy areas and time. As public input, reviews of scientific research and meetings with regulatory officials occur throughout the a rulemaking process; it is a challenge to select one document or mode of observation through which to measure influence on policymakers. RIAs offer a comparatively comprehensive and accurate measure of agency reactions to external pressure, since they are a unique expression of agencies' best arguments for proposed rules and are produced by a wide range of regulatory agencies. Finally, by often focusing on a single measure of controversy, such as public comments, previous research has failed to offer a general theory to explain the effects of political attention writ large. We improve upon this research by using measures of controversy that account for attention from three main actors in the political arena – the general public, organized interests, and the media – to develop a broad understanding of the relationships involving the use of science in policymaking.

# 3   Data and Methods

Our dataset consists of all RIAs produced for economically significant proposed rules from 2008-2012, a total of 101 documents. We draw upon the scientific citation data used in Desmarais and Hird (2013), which was gathered using RIAs from the Mercatus Center at George Mason University. Table 1 presents the breakdown of RIAs tabulated by the issuing federal agency. The RIAs are distributed across 13 agencies. About one-quarter (25) of the RIAs come from the Environmental Protection Agency, followed by the Department of Transportation with 15 RIAs and Department of Labor with 13. The top three agencies account for one-half of all RIAs produced in this time period.

[Table 1 about here.]

## 3.1   Dependent variable

The number of scientific citations in an RIA is the dependent variable. A citation was coded as "scientific" if it is published in a scholarly journal in the Institute for Scientific Information's Web of Science. This follows much bibliometric work that uses the Web of Science for information on scholarly publications (see, e.g., Braam, 2009; Vieira and Gomes, 2011; Leydesdorff, Carley and Rafols, 2013; Desmarais and Hird, 2013). See Desmarais and Hird (2013) for more information on the original data collection efforts.

The distribution of scientific citations is heavily skewed and exhibits a heavy right tail as seen in Figure 1. The number of scientific citations in a given RIA ranges from a minimum of

0 to a maximum of 258, with a median of 3 and mean of 12.2. Among the 101 RIAs, there are 1,234 total citations to scholarly articles. Given that RIAs are developed in a much different context than scientific publications – the documents in which the scientific community is accustomed to assessing the extent and quality of citations – it is difficult to say whether the levels of citation in RIAs constitutes relatively sparse or comprehensive use of science. However, the wide variation observed across RIAs can nonetheless be exploited to understand the correlates of science use within RIAs. RIA's also cite other, non-scientific, evidence supporting proposed rules, though we do not address that here.

[Figure 1 about here.]

It is important to note that we do not assess whether more science in an RIA indicates a higher quality regulation, rather only whether more attention to the proposed rule alters the volume of science utilized. Our main goal is to examine whether regulators use more science in controversial political environments, so we believe the count of scientific citations is appropriate for this purpose. A few points however should be made about this measure. First, we use the data from Desmarais and Hird (2013), where they find that articles cited in RIAs come from journals with higher impact factors than articles in the same scientific domain not cited in RIAs. While we cannot assume this translates to the quality of the regulation, the agencies in our dataset do invoke higher quality science, so there is at least some level of observable, purposeful use. Second, while articles in RIAs are published in high impact factor journals on average, there is some variation across agencies regarding the volume of scientific citations in RIAs, so high-quality science is not always invoked by all agencies. While citations can be used to either meaningfully supplement a justification with scientific evidence or merely to bolster pre-existing claims, incorporating research into RIAs is voluntary and done at the discretion of agency officials. A positive relationship between the number of scientific citations in an RIA and the political attention surrounding the rule would therefore suggest that this level of political attention influences the level, but not necessarily the type or quality, of science use.

## 3.2   Independent variables

We measure political attention to an instance of regulatory policymaking using three complementary measures: 1) public comments during the notice-and-comment period, 2) requested meetings concerning the proposed rule through OIRA, and 3) national news coverage of the legislative office (i.e., the executive branch office with legislative authority to develop the regulation). Following, Shapiro and Borie-Holtz (2014), we interpret heightened attention to a regulatory proposal as a signal of the political controversy surrounding the proposed rule. Table 2 presents descriptive statistics of all key variables in our model. Before an agency is granted authority by Congress to issue a rule, agencies can publish an Advanced Notice of Proposed Rulemaking where comments from the public are solicited. This notice-and-comment period generally lasts 60-90 days before the agency, if moving forward with the regulation, issues a proposed rule. The public then has the option to submit comments again, which the agency is required to consider before issuing a final RIA and rule. At any point in this process, interested groups (mostly industry lobbyists) can meet directly with the Office of Information and Regulatory Affairs about the proposed regulation. These meetings are generally taken to mean that these interested parties desire to change the substance of the rule; they signify controversy. Finally, media coverage can happen at any point in the rulemaking process, but we are interested in media coverage that might similarly signal increasing attention to agencies as they are developing the language of the RIA. We think that these three measures of attention – from the public, industry, and the media – that happen *before* and *during* the development of the rule will be related to the volume of science used to justify the rule.

[Table 2 about here.]

The number of public comments directed to each RIA was gathered from regulations.gov, the federal electronic docket system for rulemaking. The volume of public comments measures attention on the part of the broadest possible constituency for regulatory policymaking, and has been used in past research to assess the broad public salience of regulations (Balla and Daniels, 2007). Comments can either be submitted directly to regulations.gov or to the agency itself. Comments are filtered by agency officials before being posted to regulations.gov as documents in the document, meaning that there are often more comments submitted than are posted as documents to a docket. Regulations.gov reports that reasons underpinning agencies' filtering decisions include removing comments, "containing private or proprietary information, inappropriate language, or duplicate/near duplicate examples of a mass-mail campaign." It is not possible to know with certainty whether comments posted to a docket arose through individuals' initiatives or through coordinated campaigns. However, to address the problem of "astroturfing" in part, we use the post-filtering number of comments as our measure. Comments are most commonly submitted by the general public, but also include submissions from interest groups, members of Congress, and other organized interests. Comments are uploaded and posted to regulations.gov sometimes several months after they were originally submitted to the agency. One RIA from the Department of Labor does not have a docket page on regulations.gov and was omitted from the analysis. See Figure 2 for the comment variable's distribution.

[Figure 2 about here.]

To complement the comments measure of broad engagement, we use the number of OIRA meetings requested for specific rules to assess attention from elite policy actors. The Office of Information and Regulatory Affairs (OIRA) is required to provide meeting records for all draft regulations that require review, including meeting dates, names and affiliations of attendees, documents, and metadata about the rule. On the recently updated OIRA website, meetings are searchable by several different criteria, including Regulatory Identification Number (RIN), stage of rulemaking, date, and agency. Every proposed rule that is deemed significant by OIRA is docketed on the site with information regarding the regulation and its pending review. We used the Center for Progressive Reform's (CPR) collection of meeting records, which ranges from October 2001 to June 2011. Research assistants crosschecked a random sample of CPR's meeting data to ensure its accuracy, and then collected additional data through December 2012 to expand upon the original dataset. The RIN is used to match the RIAs in our dataset with CPR's data and to search for meeting records through OIRA's database. We measure the number of meetings temporally so that we only include those that occurred before the release date of the RIA in our dataset. The number of meetings requested through OIRA ranges from 0 to 47, with a mean of 1.46 and a median of 0.

[Figure 3 about here.]

To measure the relationship between media attention and science use, we used LexisNexis to gather the number of *Washington Post* articles written about the executive agency that issued a regulation. We take the number of *Washington Post* articles written about the RIA's agency 90 days before the release date of the RIA. Duplicate articles, indicated by a high similarity percentage by LexisNexis, are excluded from the analysis to ensure we capture unique pieces of national news coverage directed at the issuing agency. While the news articles are about the agency instead of the particular regulation itself, previous research has found that agency media attention does increase at the beginning of an issue-attention cycle, and that this increased attention leads regulators to produce immediate responses (Yanovitzky, 2002). Because media attention to regulatory policymaking is otherwise generally sparse (Gormley, 1986), our hypothesis is that a high level of media attention to the agency leading up to the finalization of a rule justification will correlate with higher RIA scientific citation counts.

[Figure 4 about here.]

## 3.3 Adjusting for Agency Science Policies

Individual agency characteristics may shape the degree to which each agency relies on scientific research in their impact analyses (Desmarais and Hird, 2013). To control for potential agency effects, we use a categorical variable that captures differences in agency communications, personnel, issue area, culture, institutional procedures, and more. In March 2009, the Obama Administration sent a memorandum to the heads of executive departments and agencies outlining broad principles to preserve and promote scientific integrity of the executive branch. In December 2010, White House Science Advisor John Holdren followed up with an executive directive providing detailed guidelines for agencies to implement the Administration's policies on scientific integrity. The directive offers information regarding the unique importance of science in rulemaking, stating that "successful application of science in public policy depends on the integrity of the scientific process both to ensure the validity of the information itself and to engender public trust in Government."(Holdren, 2010, p.1) Holdren provides specific protocols regarding public communications, the use of federal advisory committees, professional development of government scientists, and the implementation of practices that adhere to a strict policy that upholds scientific integrity.

The Union of Concerned Scientists (UCS) prepared a comparative analysis of the 22 federal agency scientific integrity policies developed in response to the memo (Grifo, 2013), including an assessment of the adequacy and quality of each. In sum, UCS writes, "six agencies submitted policies that actively promote and support a culture of scientific integrity; five submitted policies that also promote and support scientific integrity but need additional to ensure long-term change at the agencies. Eleven agencies submitted policies that do not make adequate commitments to achieve the preservation and promotion of scientific integrity."

Based on the UCS analysis, we coded agency scientific integrity on a scale from 0 to 3 where 3 indicates policies that "actively promote and support" scientific integrity, 2 indicates those that "need additional work", 1 indicates those that "do not make adequate commitments" to scientific integrity, and where 0 indicates a complete absence of a scientific integrity policy. We coded based on the closest legislative office; for example, while the Department of Health and Human Service's policy received the lowest rating by UCS, HHS agencies Food and Drug Administration (FDA) and Centers for Disease Control and Prevention (CDC) were rated higher. RIAs written by Health and Human Services, therefore, were coded as having a level 1 scientific integrity, while RIAs drafted by FDA and CDC were coded as 2 and 3, respectively.

# 4 Results

We estimate the effects of our independent variables on scientific citations using a quasi-Poisson regression because our dependent variable is a count. Quasi-Poisson regression accounts for over-dispersion in the count data and adjusts for convergence issues due to small sample sizes. For our analysis, we take the log of each independent variable (plus one) because we expect decreasing marginal effects for each additional comment, meeting, and news article, respectively. Table 3 presents the results from the quasi-Poisson model estimating scientific citations. Model 1 measures the effects of our measures of political attention and controversy on the scientific citation count in each RIA in our dataset. As expected, scientific integrity is strongly related to scientific citations; the better quality an agency's scientific integrity policy, the more it cites science. Moreover, our key independent variables of interest all positively influence the level the science invoked in RIAs. Public comments, meetings requested with OIRA officials, and news coverage all have a positive affect on the number of scientific citations, indicating that the more attention there is surrounding a proposed regulation, the more likely agency officials will invoke scientific evidence in the RIA justification. These results are robust across a wide range of model specifications.

[Table 3 about here.]

Despite within-agency consistency in the level of science citation, there is still notable variation within those agencies for which we have voluminous data in our sample. To test whether our results still hold within agencies, we run the regression only on RIAs produced by the Environmental Protection Agency, the most frequent RIA producer in our dataset. As we can see from Model 2, public comment and meetings requested with OIRA are still positively and significantly associated with scientific citations. However, once our news coverage variable is added to the model, public comments are no longer statistically significant, even though meetings and news coverage are positive indicators of citations. Although we do not find significant results across all three independent variables once we incorporate news coverage, indicating collinearity, the coefficients remain positive and comparable in magnitude to Model 1. It is therefore likely that the small sample size of 25 EPA regulatory impact analyses prevents us from finding results that are significant at conventional levels. The results presented here indicate that political attention and controversy surrounding a proposed regulation does have an overall influence on agency officials' decisions to invoke scientific evidence in their justificatory analyses. This within-agency variation suggests that individual regulatory agencies respond to external pressures, not merely that agencies issuing more controversial regulations cite science more frequently.

Since the use of science might more appropriately be thought of as the *rate* of scientific citations in an RIA, rather than the count of scientific citations, we include another model with the length of an RIA as an offset. This corrects for the number of scientific citations per page to account for an increase in the use of science as the RIA gets longer. Including an offset in a count model multiplies the baseline rate of events (i.e., citations) by the exponent of the offset. By including the natural log of page length as an offset, we multiply the baseline rate of citation by page length, which builds in the expectation that citations will scale linearly with page length. Model 4 presents the coefficients from a regression with page length included on the right-hand side as an offset on the log scale. With this offset included, meetings requested with OIRA officials as well as recent news attention both significantly predict the use of science. Public comments and scientific integrity policies, however, are not statistically significant in the model with the page length offset.

[Figure 5 about here.]

Due to the nonlinear and multiplicative nature of predicted values using quasi-Poisson regression, additional manipulation is necessary to substantively interpret our results. Figure 5 presents predicted scientific citations based on the level of scientific integrity in the agency's internal policy using the effects from Model 1. An agency with the highest rated scientific integrity policy has a predicted scientific citation count of almost 10 citations, while the predicted value of scientific citations for an agency with no integrity would be just 3, which is the baseline value predicted by setting all other variables to their medians.

[Figure 6 about here.]

[Figure 7 about here.]

Figures 6 and 7 present the predicted number of scientific citations in RIAs based on both the number of OIRA meetings addressing the regulation and the number of comments on the regulation's docket. Each additional OIRA meeting leads to approximately one additional predicted cite. Comments also exhibit a substantively significant relationship with the predicted number of citations. A one standard deviation increase in the number of comments leads to nearly a 50% increase in predicted scientific citations.

In regards to the meetings variable, the CPR data on meeting records from OMB includes the number of each kind of meeting participant, such as industry, White House officials, firms, interest groups, and so on. We also analyzed the relationship between the type of participant and the number of scientific citations in order to examine whether one type of meeting participant, such as industry-related attendees, were driving the number of meeting effects. However, we found no significant correlation between meeting participant type and scientific citation counts, suggesting that the effects of OIRA meetings are indeed a result of the sheer number of meetings rather than the types of groups requesting and attending them.

Finally, Figure 8 presents the predicted number of citations based on news coverage, measured by *Washington Post* news articles about the issuing agency published during the production of a regulatory impact analysis. There is again a significantly positive relationship between news coverage and scientific citations, where each additional news article results in approximately 0.5 additional cites.

[Figure 8 about here.]

## 5 Conclusion

Scientific research is regularly cited in the regulatory impact analyses produced in support of major regulations. In this paper we examine whether science is used more heavily in highly contested regulatory contexts, which has not been previously tested empirically by scholars of public policy or bureaucratic behavior. While scholars have argued controversy inflates the need for research, their focus has varied from administrative law (Melnick 1983, 1992; Jasanoff 1990) to specific policy outcomes or arenas (for example, Contandriopoulos et al., 2010; Farrow, 1991; Whiteman, 1985). Overall, we find support for the hypothesis that more controversy surrounding a proposed rule increases the use of science in rule justifications.

The implications of our findings are twofold. First, by offering the first analysis of science use in RIAs across these measures of controversy, our study sets the groundwork for examining connections between external pressures of the political environment and science use in regulatory policymaking across time and regulatory domains. Our findings show a significant correlation between political attention – from the general public, organized interests, and the media – and the invocation of science in justificatory regulatory policy documents. Future research can build upon this study by examining these effects in greater detail. Both key independent variables, public comments and OIRA meetings, should be explored further to examine more nuanced relationships between organized interests and regulators. For example, while the volume of comments might signal to the agency a high level of political salience, it is also the content of the comments that might affect the agency's use of knowledge; that is, if commenters link to scientific literature or provide technical information, agency officials may take that into consideration and invoke more science when writing the RIA. However, in a study of business interest comments on regulations, Yackee and Yackee (2006) found no significant correlation between comment quality (technical or expert information provided) and the extent of change in the final rule, indicating that the type of commenter (business or not) is more important to the agency than the expertise level of the commenter. It is possible that only after a certain threshold of both comments and commenter-type do agencies start paying significant attention to the expertise and technical information provided. Additionally, it is possible that agencies address the scientific information in the RIA without significantly altering the final rule. A more in-depth study involving content analysis of individual comments would help to answer these questions. In regards to meetings requested through OIRA, analyzing in greater depth the actors involved and the documents or evidence presented at the meetings could not only shed light on our understanding of the OIRA review and meeting process, but in vested interests to regulatory policymaking in general.

More broadly, our findings have implications for participatory democracy and rulemaking. Agencies are responsive to their political environment and pressures of external attention, and are able to invoke more scientific evidence when deemed necessary. High-stakes, controversial rules are more likely to compel regulators to invoke science in their public justifications. To be sure, this does not mean that agencies otherwise shirk their responsibility to use good science for all regulatory decisions. Rather, one external condition that correlates with the volume of science used is political controversy. Based on our findings, science plays a more crucial role when controversy is signaled through the general public and vested interests. Since agencies include citations in RIAs on their own initiative, scientific citations are not mere window dressing on regulations but appear to be a purposeful use of evidence to bolster arguments supporting regulatory proposals. Indeed, if controversy is associated with the impact of the rules involved, greater use of science in highly salient rulemaking is an appropriate allocation of scarce bureaucratic resources; important rules require greater use of science. Based on our finding that agencies systematically invoke more science with increasing levels of controversy, it appears that science plays a more crucial role in the regulatory domains of greatest interest to the public and other policy actors.